\documentclass[11pt]{article}
\pdfoutput=1

\usepackage{euscript}
\usepackage{amssymb}
\usepackage{amsfonts}
\usepackage{amsbsy}
\usepackage{epsfig}
\usepackage{amsthm}
\usepackage{amscd}
\usepackage{amstext}
\usepackage{verbatim}
\usepackage{amsmath}
\usepackage{cancel}
\usepackage{capt-of}
\usepackage{empheq}
\usepackage{subfigure}
\usepackage{xcolor}

\usepackage[pdftex]{hyperref}
\hypersetup{ 
colorlinks=true, 
linkcolor=black, 
citecolor=red, 
}

\def\ben{\begin{equation}}
\def\een{\end{equation}}

   \let\d=\delta

\let\w=\omega

\let\pa=\partial
\def\be{\begin{equation}}
\def\ee{\end{equation}}
\def\beq{\begin{equation}}
\def\eeq{\end{equation}}
\def\ba{\begin{array}}
\def\ea{\end{array}}

\def\dalemb#1#2{{\vbox{\hrule height .#2pt
       \hbox{\vrule width.#2pt height#1pt \kern#1pt
               \vrule width.#2pt}
       \hrule height.#2pt}}}

\newcommand{\bea}{\begin{eqnarray}}
\newcommand{\eea}{\end{eqnarray}}

\renewcommand{\eqref}[1]{(\ref{eq:#1})}

\def\ocal{{\mathcal{O}}}

\def\hri#1#2{\href{http://arxiv.org/abs/#1}{[ArXiv:#1]#2}}
\def\hre#1#2{\href{http://arxiv.org/abs/#1/#2}{[ArXiv:#1/#2]}}

\begin{document}
\pagenumbering{gobble}

\begin{flushright}
SU-ITP-15/11
\end{flushright}
\begin{center}

{ \Large {\bf
Incoherent transport in clean quantum critical metals
}}

\vspace{1cm}

Richard A. Davison$^{1}$, Blaise Gout\'eraux$^{2,3}$ and Sean A. Hartnoll$^{2}$

\vspace{1cm}

{\small
$^{1}${\it Lorentz Institute for Theoretical Physics, \\ Niels Bohrweg 2, Leiden NL-2333 CA, The Netherlands }}

\vspace{0.5cm}

{\small
$^{2}${\it Department of Physics, Stanford University, \\
Stanford, CA 94305-4060, USA }}

\vspace{0.5cm}

{\small
$^{3}${\it APC, Universit\'e Paris 7, CNRS, CEA, \\ Observatoire de Paris, Sorbonne Paris Cit\'e, F-75205, Paris Cedex 13, France}}

\vspace{1.6cm}

\end{center}

\begin{abstract}

In a clean quantum critical metal, and in the absence of umklapp, most d.c.~conductivities are formally infinite due to momentum conservation. However, there is a particular combination of the charge and heat currents which has a finite, universal conductivity. In this paper, we describe the physics of this conductivity $\sigma_Q$ in quantum critical metals obtained by charge doping a strongly interacting conformal field theory. We show that it satisfies an Einstein relation and controls the diffusivity of a conserved charge in the metal. We compute $\sigma_Q$ in a class of theories with holographic gravitational duals. Finally, we show how the temperature scaling of $\sigma_Q$ depends on certain critical exponents characterizing the
quantum critical metal. The holographic results are found to be reproduced by the scaling analysis, with the charge density operator becoming marginal in the emergent low energy quantum critical theory.

\end{abstract}

\pagebreak
\pagenumbering{arabic}
\tableofcontents

\section{Introduction}

Condensed matter systems tuned to quantum critical points can often exhibit universal quantum physics \cite{subirbook}.
For example, dissipative processes involving the collective low energy degrees
of freedom are robustly described within an effective quantum critical theory.
Linear response conductivities at low temperatures and frequencies are prototypical instances of low energy dissipative observables. The Kubo formula, in particular, gives the d.c.~conductivities in terms of the low energy spectral weight of conserved densities \cite{mahan}. If the d.c.~electrical conductivity of quantum critical metals could be successfully related to a scale invariant quantum field theory, it would potentially explain intriguing similarities in transport observed across a range of quantum critical systems \cite{keimersachdev,andymac}.

There is, however, an interesting obstruction to relating conductivities directly to universal dissipative dynamics.
If there is a conserved quantity (typically the momentum) that overlaps with any currents of the conserved charges,
then the corresponding d.c.~conductivities are infinite. A simple example of this phenomenon, relevant to our discussion below, is transport in a conformal field theory (CFT) with a conserved electric charge (but in a state with zero charge density). Such theories arise at quantum critical points such as the Bose-Hubbard model at integer filling \cite{subirbook}. The conserved quantities are the total energy, total charge and total momentum. The corresponding currents are the heat current $J^Q$ (equal to the momentum in a CFT because $T^{ti} = T^{it}$), the electrical current $J$ and the momentum current $J^P$. Hydrodynamic arguments give the low frequency conductivities
\begin{equation}
\label{eq:neutralhydroconductivs}
\sigma_{J^Q J^Q}(\omega)=sT\left(\delta(\omega)+\frac{i}{\omega}\right),\;\;\;\;\;\;\;\;\;\;\;\;\;\sigma_{J J}(\omega)=\sigma_Q,\;\;\;\;\;\;\;\;\;\;\;\;\;\sigma_{J^P J^P}(\omega)=\eta \, .
\end{equation}
Here $s$ is the entropy density of the system and $T$ is the temperature. The charge and momentum conductivities are given by the first order hydrodynamic transport coefficients $\sigma_Q$ (the universal conductivity) and $\eta$ (the shear viscosity) of the state. The thermal conductivity, in contrast, is infinite. To render this conductivity finite, momentum 
must relax. Under weak translational symmetry breaking, $\delta(\omega)+i/\omega \to 1/(-i \omega + \Gamma)$.
The momentum relaxation rate $\Gamma$ is then determined by irrelevant (in the renormalization group sense)
corrections to the universal low energy physics. While this situation leads to an elegant description of transport that can be of practical use, see for instance \cite{Hartnoll:2007ih, Hartnoll:2012rj, Hartnoll:2014gba, Patel:2014jfa, Lucas:2015pxa}, it also
means that that the d.c.~thermal conductivity is not a fully universal quantity.\footnote{An interesting exception to this statement arises when the weak momentum relaxing processes are themselves captured in the low energy hydrodynamic theory \cite{spivak1, spivak2, Davison:2013txa, Balasubramanian:2013yqa}. In such cases the momentum relaxation rate is determined by the transport coefficients, including the viscosity.} One way to recover universal conductivities is to consider circumstances where momentum
conservation is very strongly broken in the low energy theory, as in these cases the conductivity is captured through `incoherent' universal diffusive dynamics \cite{Hartnoll:2014lpa, Davison:2014lua, Lucas:2015lna}.

The starting point of this paper will be the observation \cite{Davison:2015bea} that even in clean systems with a fully conserved momentum and a nonzero charge density (i.e. in `metallic' systems), there is a diffusive mode that can be decoupled from the conserved momentum (which is associated to sound modes). This mode has a corresponding finite d.c.~conductivity that can be computed entirely within the universal low energy physics. In equation (\ref{eq:neutralhydroconductivs}) the universal conductivities are decoupled from the non-universal thermal conductivity. For a CFT deformed by a nonzero charge density $\rho$, the decoupling is not so immediate. Hydrodynamic arguments give the thermoelectric conductivities \cite{Hartnoll:2007ih}
\begin{equation}
\label{eq:finitedensityconducs}
\begin{aligned}
\sigma_{J^Q J^Q}(\omega)&=\frac{s^2T^2}{\epsilon+P}\left(\delta(\omega)+\frac{i}{\omega}\right)+\mu^2\sigma_Q \,, \\
\sigma_{JJ}(\omega)& =\frac{\rho^2}{\epsilon+P}\left(\delta(\omega)+\frac{i}{\omega}\right)+\sigma_Q \,,  \\
\sigma_{JJ^Q}(\omega) =\sigma_{J^Q J}(\omega) & =\frac{\rho s T}{\epsilon+P}\left(\delta(\omega)+\frac{i}{\omega}\right) - \mu\sigma_Q \,, \\
\end{aligned}
\end{equation}
while the momentum conductivity is unchanged from (\ref{eq:neutralhydroconductivs}). Here $\epsilon$ and $P$ are the energy density and pressure of the state respectively and $\mu$ the chemical potential. The heat current is now $J^{Q\,i} = T^{ti} - \mu J^i$. The first few observations in this paper will be that (i) the `incoherent current'
\be\label{eq:jinc}
J^\text{inc} \equiv \frac{s T \, J - \rho \, J^Q}{\epsilon + P} \,,
\ee
carries no momentum and therefore (ii) from the equations (\ref{eq:finitedensityconducs}) together with the thermodynamic identity
$\epsilon + P = sT + \mu \rho$, has a universal d.c.~conductivity given by
\be\label{eq:sQ}
\sigma_{J^\text{inc} J^\text{inc}}(\omega) = \sigma_Q \,.
\ee
In fact, we will see shortly that (iii) this current is associated to a  conserved density that obeys a decoupled diffusion equation and $\sigma_Q$ consequently satisfies an Einstein relation. In section \ref{sec:sigma} we obtain this universal conductivity explicitly in certain examples that are described by holographic duality \cite{Maldacena:1997re, Hartnoll:2011fn} and then give a general discussion of the temperature scaling of $\sigma_Q$ in section \ref{sec:scaling}. 

Although we will mainly concentrate on the case of a CFT deformed by a nonzero charge density, many of our observations are applicable more generally. In a system whose only conserved vectorial quantity is the total momentum $P$, there is an incoherent current
\begin{equation}
\label{eq:generalincoherentcurrent}
J^\text{inc}=J-\frac{\chi_{JP}}{\chi_{PP}}P,
\end{equation}
where $\chi$ denotes the static susceptibility. By construction, this current does not overlap with the momentum ($\chi_{J^\text{inc}P}=0$), and so has a finite conductivity. In the case of a CFT at nonzero charge density, the current in (\ref{eq:generalincoherentcurrent}) is equal to that in (\ref{eq:jinc}).
 
A work very much in the same spirit as this one is \cite{Sonner:2013aua}, which studied universal transport in bilayer metals with two conserved charges. In that case a current operator carrying no momentum can also be constructed.

\section{Diffusion in CFT hydrodynamics}

The framework of our discussion will be the hydrodynamics of a CFT. Hydrodynamics is the effective theory describing the long wavelength and small frequency properties of a state near thermal equilibrium. The basic equations of relativistic hydrodynamics are firstly the conservation laws for the energy momentum tensor $T^{\mu\nu}$ and $U(1)$ current $J^\mu$:
\be
\pa_\mu T^{\mu\nu} = 0 \,, \qquad \pa_\mu J^\mu = 0 \,.
\ee
Secondly, the constitutive relations for parity-invariant relativistic hydrodynamics to first order in derivatives, and in Landau frame, are \cite{Kovtun:2012rj}
\begin{equation}
\begin{aligned}
T^{\mu\nu}&=\epsilon u^\mu u^\nu+P\Delta^{\mu\nu}-\eta\Delta^{\mu\alpha}\Delta^{\nu\beta}\left(\partial_\alpha u_\beta+\partial_\beta u_\alpha-\frac{2}{d}\eta_{\alpha\beta}\partial_\lambda u^\lambda\right)-\zeta\Delta^{\mu\nu}\partial_\lambda u^\lambda+\ldots,\\
J^\mu&= \rho u^\mu-\sigma_QT\Delta^{\mu\nu}\partial_\nu\left(\frac{\mu}{T}\right)+\ldots, \label{eq:Jmu}
\end{aligned}
\end{equation}
where $\Delta^{\mu\nu}=\eta^{\mu\nu}+u^\mu u^\nu$ is the projector, $\epsilon$ is the energy density, $P$ is the pressure, $\rho$ is the charge density, and $d$ is the number of spatial dimensions. There are three first order dissipative transport coefficients: shear viscosity $\eta$, bulk viscosity $\zeta$ and `conductivity' $\sigma_Q$. The constitutive relations above have been constrained by Lorentz invariance, but we have not yet imposed the constraint of scale invariance.

Before proceeding to solve these equations, we should note that there are two senses in which these equations can describe quantum critical systems. Firstly, we can think of the CFT itself describing the dynamics of a quantum critical point, such as the superfluid-insulator transition in the Bose-Hubbard model at integer filling \cite{subirbook}. The charge density $\rho$ appearing in (\ref{eq:Jmu}) then corresponds to a deformation away from the quantum critical point. A second perspective is that the CFT is not itself the system of primary interest but is a useful starting point to construct quantum critical finite density systems or more generally `compressible phases' in the language of \cite{Sachdev:2012dq, Sachdev:2012tj}. A weakly coupled example would be doping graphene away from its particle-hole symmetric point. If the doped CFT flows at low energies to a new finite density fixed point (that will typically not be Lorentz invariant), then the above equations of hydrodynamics will still apply except that now both the thermodynamics and transport coefficients ($\eta, \zeta, \sigma_Q$) will be properties of the low energy quantum critical metal, not the original CFT (note however that the high energy CFT implies that the bulk viscosity $\zeta = 0$ at all scales).

If the conserved charges are perturbed away from equilibrium by $\{\delta T^{tt}, \delta T^{tx}, \delta J^t\}$, standard hydrodynamic manipulations give the following equations of motion in the longitudinal channel \cite{Kovtun:2012rj}
\begin{equation}
\label{eq:hydroeqsofmotion}
\partial_t\begin{pmatrix}\delta T^{tt}\\ \delta T^{tx} \\ \delta J^t\end{pmatrix}+\begin{pmatrix}0&ik_x&0\\ik_x\beta_1&\gamma_sk_x^2&ik_x\beta_2\\\sigma_Q\alpha_1k_x^2&ik_x\frac{\rho}{\epsilon+P}&\sigma_Q\alpha_2k_x^2\end{pmatrix}\begin{pmatrix}\delta T^{tt}\\ \delta T^{tx} \\ \delta J^t\end{pmatrix}=0,
\end{equation}
where the various thermodynamic quantities are
\begin{equation}
\begin{aligned}
\label{eq:alphabetadefns}
\alpha_1&=\left(\frac{\partial\mu}{\partial\epsilon}\right)_\rho-\frac{\mu}{T}\left(\frac{\partial T}{\partial\epsilon}\right)_\rho,\qquad \beta_1=\left(\frac{\partial P}{\partial\epsilon}\right)_\rho \,, \\
\alpha_2 &=\left(\frac{\partial\mu}{\partial \rho}\right)_\epsilon-\frac{\mu}{T}\left(\frac{\partial T}{\partial \rho}\right)_\epsilon\,, \qquad\beta_2=\left(\frac{\partial P}{\partial \rho}\right)_\epsilon \,,
\end{aligned}
\end{equation}
and
\begin{equation}
\gamma_s=\frac{\frac{2d-2}{d}\eta+\zeta}{\epsilon+P} \,.
\end{equation}

The solutions of the above equations give the coupled hydrodynamic modes. Let us now decouple the diffusive mode. It is straightforward to verify that the linear combination
\begin{equation}
\label{eq:perturbativediffusionoperator}
\delta Q^{\text{diff}}=\delta J^t-\frac{\rho}{\epsilon + P}\delta T^{tt}+\sigma_Q\frac{\alpha_1+\frac{\rho}{\epsilon + P}\alpha_2}{\beta_1+\frac{\rho}{\epsilon + P}\beta_2}\partial_i \delta T^{ti}+O(\partial^2) \,,
\end{equation}
obeys the diffusion equation
\begin{equation}
\label{eq:decouplediffusioneq}
\partial_t \delta Q^{\text{diff}}+Dk^2 \delta Q^{\text{diff}}+O(k^3)=0 \,,
\end{equation}
where the diffusion constant is (in agreement with \cite{Kovtun:2012rj})
\begin{equation}
\label{eq:perturbativediffusioneq}
D=\sigma_Q\frac{\alpha_2\beta_1-\alpha_1\beta_2}{\beta_1+\frac{\rho}{\epsilon + P}\beta_2}\,.
\end{equation}
Here we have worked to the order in wavevector $k$ consistent with the first order hydrodynamic constitutive relations above.

In (\ref{eq:perturbativediffusionoperator}) we see that the diffusing quantity involves not only the original fluctuations of the conserved densities, but also derivatives of these densities. However, so far we have not used conformal invariance. An interesting simplification occurs in this case. Scale invariance implies that the equation of state is $\epsilon = d \, P$ and hence in (\ref{eq:alphabetadefns}) we have $\beta_2 = 0$. Employing further thermodynamic manipulations on (\ref{eq:alphabetadefns}):
\be
\frac{\alpha_2}{\alpha_1} = - \frac{T\left(\frac{\partial\epsilon}{\partial T}\right)_\mu+\mu\left(\frac{\partial\epsilon}{\partial\mu}\right)_T}{\left(\frac{\partial\epsilon}{\partial\mu}\right)_T} = - \frac{\epsilon + P}{\rho} \,.
\ee
It follows that in conformal relativistic hydrodynamics, the incoherent charge density (\ref{eq:perturbativediffusionoperator}) becomes simply (writing $\delta J^t = \delta \rho$ and $\delta T^{tt} = \delta \epsilon$)
\be\label{eq:qdiff}
\delta Q^{\text{diff}} = \delta \rho -\frac{\rho}{\epsilon + P}\delta \epsilon = \frac{s T \, \delta \rho - \rho \, T\delta s}{\epsilon + P} 
= \frac{T s^2}{\epsilon + P} \delta \left(\frac{\rho}{s} \right) \,.
\ee
For the second equality we used the thermodynamic identity $\epsilon + P = sT + \mu \rho$ as well as the first law $\delta \epsilon = T \delta s + \mu \delta \rho$. Equation (\ref{eq:qdiff}) makes explicit that this combination of charge densities indeed corresponds to the incoherent current in (\ref{eq:jinc}), so that
\be
\frac{\partial}{\partial t} \delta Q^{\text{diff}} + \nabla \cdot J^\text{inc} = 0 \,.
\ee
Recall here that the entropy current is the heat current $J^Q$ divided by the temperature. Thus we see that the incoherent conductivity $\sigma_Q$ in (\ref{eq:sQ}) is indeed the universal conductivity associated with diffusion of a conserved charge
$\delta Q^{\text{diff}}$. More generally, the standard manipulations of Kadanoff and Martin \cite{km,Kovtun:2012rj} now imply that the hydrodynamic retarded Green's function for the current will be
\be
G^R_{J_x^\text{inc}J_x^\text{inc}} = \frac{\omega^2 \sigma_Q}{-i \omega + D k^2} \,.
\ee
The conductivity is obtained from the Green's function for the current by the usual expression $\sigma_\text{inc}(\omega) = G^R_{J^\text{inc}J^\text{inc}}(\omega,0)/(i \omega)$.

The diffusivity (\ref{eq:perturbativediffusioneq}) can be written -- even without assuming conformal invariance -- in the form of an Einstein relation
\be
D = \frac{\sigma_Q}{\chi^{\text{inc}}} \,.
\ee
Here the susceptibility $\chi^{\text{inc}}$ for the incoherent density fluctuation $\delta Q^{\text{diff}}$ is
\be\label{eq:chiinc}
\chi^{\text{inc}} = \chi_{\d Q^{\text{diff}} \d Q^{\text{diff}}} = \chi_{\delta \rho \, \delta \rho} - \frac{2 \rho}{\epsilon + P} \chi_{\delta \rho \, \delta \epsilon} + \frac{\rho^2}{(\epsilon + P)^2} \chi_{\delta \epsilon \, \delta \epsilon} \,.
\ee
The derivation of (\ref{eq:chiinc})  from (\ref{eq:perturbativediffusioneq}) using thermodynamic identities can be found in \cite{Kovtun:2012rj}. Our contribution here is to emphasize that the diffusive mode should be understood as transporting a conserved density whose
current carries no momentum. We have seen that this interpretation is especially crisp in the presence of conformal invariance.

Having clarified the physics of $\sigma_Q$, we proceed to obtain $\sigma_Q$ in a few examples. These will be cases in which the strongly interacting CFT doped to a nonzero charge density admits a dual holographic gravity description. It should be emphasized again that $\sigma_Q$ is not a property of the particle-hole symmetric CFT. Instead, it is a property of the low energy compressible phase to which the CFT flows upon deformation by a chemical potential.

\section{Holographic formula for \texorpdfstring{$\sigma_Q$}{$sigmaQ$}}
\label{sec:sigma}

From (\ref{eq:finitedensityconducs}) we see that a clean way to obtain $\sigma_Q$ is to compute
\be
\sigma_Q = \lim_{\omega \to 0} \frac{\text{Im} \, G^R_{J^x J^x}(\omega)}{\omega} \,.
\ee
Note that this formula does not care about any delta function that may be present at $\omega = 0$.
A rather general holographic formula for this quantity can be obtained.
In holographic duality, the retarded Green's function of a current $J^x$ is obtained by solving the dual bulk Maxwell equations for perturbations of the bulk field $a_x$ about a background spacetime \cite{Hartnoll:2009sz}. Both the background and the fluctuation equations must come from some bulk action. A broad class of holographic actions take the form of Einstein-Maxwell theory coupled to matter fields, which at this point we can allow to be charged and to couple non minimally to the Maxwell field:
\be\label{eq:action}
{\mathcal L} = R - \frac{Z_\text{mat.}}{4} F_{\mu\nu}F^{\mu\nu} + {\mathcal L}_\text{mat.}
\ee
Here, both $Z_\text{mat.}$ and ${\mathcal L}_\text{mat.}$ are functions of the matter fields. We will obtain a formula for $\sigma_Q$ in this class of theories.

For translationally invariant, isotropic solutions of the equations of motion, the background metric and electrostatic potential take the form
\be
ds^2 = -D(r) dt^2 + B(r) dr^2 + C(r) \left( dx^2 + dy^2 \right) \,, \quad A_t(r) = A(r) \,.
\ee
We will specialize to $d=2$ boundary spatial dimension in the explicit holographic computations, although we will give results for general $d$. All that is required of the remaining fields in the action is that they be functions of $r$ only. In particular, evaluated on the solution
\be
Z_\text{mat.} = Z(r) \,.
\ee
For the first part of the computation we do not need to know anything about the matter Lagrangian ${\mathcal L}_\text{mat.}$
beyond the fact that it does not depend on derivatives of the Maxwell field. A non-derivative dependence on the Maxwell field itself is allowed and means that this part of the analysis applies to cases with charge outside the horizon, such as holographic superconductors and electron stars \cite{Hartnoll:2011fn}. With the above assumptions, the linearized equation of motion for a spatially homogeneous ($k = 0$) perturbation $a_x(t,r)$ about the background takes the form
\be\label{eq:ax}
\frac{1}{\sqrt{BD}} \left(\sqrt{\frac{D}{B}} Z a_x' \right)'  - \frac{Z}{D} \frac{\pa^2 a_x}{\pa t^2} = \left(\frac{Z^2 A'^2}{B D} + \cdots \right) a_x \,.
\ee
Here the $\cdots$ terms depend on the possible mass terms for the vector potential due to screening by charged matter in the bulk.

All we need to know about the background metric at this point is that asymptotically, as $r \to \infty$, it tends to $AdS_4$, i.e.
\be\label{eq:glarger}
ds^2 \to - r^2 dt^2 + \frac{dr^2}{r^2} + r^2 \left(dx^2 + dy^2 \right) \,.
\ee
The metric must furthermore have a regular horizon as $r \to r_+$, so that
\be\label{eq:gnearh}
ds^2 \to - 4 \pi T (r-r_+) dt^2 + \frac{dr^2}{4 \pi T (r-r_+)} + \frac{s}{4 \pi} \left(dx^2 + dy^2 \right) \,.
\ee
Furthermore we take $Z \to 1$ at the asymptotic boundary (this amounts to choosing the normalization of charge) and $Z \to Z_+$ at the horizon.

\subsection{General formula in terms of horizon data}

With the above assumptions at hand, we can follow the elegant argumentation in \cite{Lucas:2015vna} to obtain a formula for $\sigma_Q$ in terms of horizon data of the equation (\ref{eq:ax}) with no time dependence, $\pa_t = 0$. Firstly, let $a_x^{(0)}(r)$ be the time-independent solution of (\ref{eq:ax}) that tends to one at the asymptotic boundary and which is regular on the horizon. The second solution is then, using the Wronskian method to find the second solution of (\ref{eq:ax}) in terms of the first,
\be\label{eq:second}
a_x^{(1)}(r) = a_x^{(0)}(r) \int_r^\infty \left[\sqrt{\frac{D}{B}} Z \left( a_x^{(0)}\right)^2 \right]^{-1} dr \,.
\ee
As $r \to \infty$, using the asymptotic form (\ref{eq:glarger}) of the metric, we have
\be
a_x^{(1)}(r) \to \frac{1}{r} \,.
\ee
At the horizon $r \to r_+$, from the near horizon form (\ref{eq:gnearh}) of the metric we have
\be\label{eq:ax11}
a_x^{(1)}(r) \to - \frac{1}{Z_+ a_x^{(0)}(r_+) 4 \pi T} \log \left( r - r_+ \right) + \text{finite} \,. 
\ee

Near the horizon, the solution to (\ref{eq:ax}) must satisfy infalling boundary conditions \cite{Son:2002sd,Hartnoll:2009sz}. Writing $a_x(t,r) = a_x(r) e^{- i \omega t}$, this means that to leading order near the horizon \cite{Hartnoll:2009sz}
\be
a_x(r) = a_x^{(0)}(r_+) \, e^{- \frac{i \omega}{4 \pi T}  \log \left( r - r_+ \right)}  + \cdots \,.
\ee
Moving a little away from the horizon and then expanding to first order in small $\omega$ gives
\be
a_x(r) = a_x^{(0)}(r_+) \left(1 - \frac{i \omega}{4 \pi T}  \log \left( r - r_+ \right) \right) + \cdots \,.
\ee
The $\cdots$ terms here include order $\omega$ terms coming from the finite part of (\ref{eq:ax11}). However, these will be real and will not contribute to the imaginary part of the Green's function that we are after.
Recalling the form of the second solution (\ref{eq:second}), and only worrying about the imaginary part of the response, it follows that the full solution everywhere (except right at the horizon) to first order in $\omega$ must be
\be
a_x(r) = a_x^{(0)}(r) + i \omega \, Z_+ \left( a_x^{(0)}(r_+) \right)^2  a_x^{(1)}(r) \,,
\ee
so that expanding near the boundary
\be
a_x(r) \to 1 + i \omega Z_+ \left( a_x^{(0)}(r_+) \right)^2 \frac{1}{r} \,.
\ee
From the usual AdS/CFT dictionary \cite{Hartnoll:2009sz}, the Green's function is the ratio of the normalizable by the non-normalizable mode near the boundary, so that
\be\label{eq:sigmaQ1}
\sigma_Q = Z_+ \left( a_x^{(0)}(r_+) \right)^2 \,.
\ee
Many previous works on applied holography have expressed the imaginary part of the retarded Green's function in terms of quantities evaluated on the horizon (e.g. \cite{Son:2002sd} is an early instance). The argument we have followed here (from \cite{Lucas:2015vna}) is rather tidy and does not explicitly use the action. It holds with or without charge outside the horizon.

\subsection{Explicit formula for a massless bulk photon}

In order to use the horizon formula (\ref{eq:sigmaQ1}) to get an explicit formula for $\sigma_Q$, it
is necessary to solve the perturbation equation (\ref{eq:ax}) in the time-independent case.
This equation can be solved quite elegantly in the case where the $\cdots$ terms in (\ref{eq:ax}) are absent.
That is, when there is no charged matter in the bulk. Any form of neutral matter is otherwise allowed in the
bulk so long as it does not couple directly to the Maxwell field, except through $Z$.

The important step for solving the equation is to express it as a total derivative. This trick goes back to the
work of Iqbal and Liu \cite{Iqbal:2008by} for the case with no charge density ($\rho = 0$). The argument we are about to give generalizing that result to the action (\ref{eq:action}) and finite charge density states (with $\rho \neq 0$) has already appeared in the very nice papers \cite{Jain:2010ip, Chakrabarti:2010xy}. Our presentation is perhaps slightly more streamlined, but it amounts to the same derivation.

To simplify the perturbation equation (\ref{eq:ax}) we need two equations for the background functions that hold independently of the matter fields, so long as the matter fields are not charged. Firstly, the Maxwell equation is
\be\label{eq:first}
\frac{d}{dr} \left( \frac{C Z}{\sqrt{BD}} A' \right) = 0  \qquad \Rightarrow \qquad \frac{C Z}{\sqrt{BD}} A' = \rho \,.
\ee
To evaluate the constant in the second equation we used the asymptotic expansion of the metric functions (\ref{eq:glarger})
as well as the fact that, as per the standard holographic dictionary \cite{Hartnoll:2009sz}, as $r \to \infty$ the Maxwell field behaves as $A(r) \to \mu - \rho/r$.

The second equation needed is that 
\bea
\lefteqn{\frac{d}{dr} \left( \frac{C}{\sqrt{BD}} \left(Z A A' - C \left(\frac{D}{C}\right)' \right) \right) = 0} \nonumber \\
&& \qquad \Rightarrow \qquad \frac{C}{\sqrt{BD}} \left(Z A A' - C \left(\frac{D}{C}\right)' \right) = - s T \,. \label{eq:seconds}
\eea
The second line comes from evaluating the constant on the horizon, as $r \to r_+$. We used the near horizon form of the metric (\ref{eq:gnearh}) together with the fact that $A(r_+) = 0$ for regularity of the Euclidean solution. The equation in the first line can be obtained as the conservation of a Noether charge of a certain scaling symmetry of the action (\ref{eq:action}) on radially dependent solutions, see \cite{Gubser:2009cg}. The Noether symmetry argument goes through in the presence of matter fields because the scaling symmetry leaves $\sqrt{-g}$, $g_{rr}$ and $g^{tt} A_t^2$ invariant. For configurations where the fields only depend on $r$, only these three combinations of the metric and Maxwell field appear in the matter action (allowing even for charged fields) and hence the matter fields themselves do not transform under the symmetry. Additional vector fields however would lead to additional terms in (\ref{eq:seconds}).

Using the two equations just derived, the zero frequency perturbation equation (\ref{eq:ax}) for $a_x$ can be written in the `massless' form
\be\label{eq:EMD}
\frac{d}{dr} \left(\frac{C^2 Z}{B} \left(\frac{D}{C}\right)' a_x' - \frac{C Z^2 A'^2}{B} a_x \right) = 0 \,.
\ee
This radially conserved quantity has an immediate physical meaning. It asymptotes to the incoherent current $J^\textrm{inc}$ at the boundary $r\to+\infty$.\footnote{This fact provides a distinct avenue to calculating $\sigma_Q$ than the one we are following here. One can use the conserved quantity in (\ref{eq:EMD}) to obtain directly the d.c.~limit of $\sigma_{\text{inc}}(\omega)$ using the method of \cite{Donos:2014cya}.} From the fact that $B$ diverges at the horizon, we see that the solution that is regular at the horizon must in fact obey
\be
C \left(\frac{D}{C}\right)' a_x' - Z A'^2 a_x = 0 \,.
\ee
This equation is immediately integrated. Imposing that $a_x$ go to one at the asymptotic boundary gives the solution
\be\label{eq:axo}
a_x^{(0)} = \exp\left\{ - \int^\infty_r \frac{Z A'^2}{C (D/C)'} dr \right\} \,.
\ee
Simple algebra using the two constants of motion (\ref{eq:first}) and (\ref{eq:seconds}) shows that the integral in the exponent of (\ref{eq:axo}) is
\be
\int^\infty_{r} \frac{Z A'^2}{C (D/C)'} dr = \int^\infty_{r} \frac{A'}{A + sT/\rho} dr = \log \frac{\epsilon + P}{A(r) \rho + sT} \,.
\ee
For the last equality we used the fact that $A(\infty) = \mu$ and that $\epsilon + P = sT + \mu \rho$. This second statement is the 
Smarr law and can be obtained from (\ref{eq:seconds}) by evaluating the constant at $r \to \infty$, and extracting the energy density $\epsilon$ and pressure $P$ from the normalizable falloffs of the metric, in the standard way \cite{Hartnoll:2009sz}.
Therefore from (\ref{eq:axo}) 
\be\label{eq:aox}
a_x^{(0)}(r) = \frac{A(r) \rho + sT}{\epsilon + P} \,.
\ee
Recall again that at the horizon $A(r_+) = 0$. Thus
\be
a_x^{(0)}(r_+) = \frac{sT}{\epsilon + P} \,.
\ee
It now follows, using the previous result (\ref{eq:sigmaQ1}), that
\be
\sigma_Q = Z_+ \left( \frac{sT}{\epsilon + P} \right)^2 \,.
\ee
This is the result obtained in \cite{Jain:2010ip, Chakrabarti:2010xy} (and earlier in \cite{Hartnoll:2007ip} for the case of Einstein-Maxwell theory). The upshot is that the incoherent conductivity $\sigma_Q$ is given, in these theories, by thermodynamic quantities multiplied by a certain `horizon conductivity' $Z_+$ that appears in the membrane paradigm description of horizons \cite{Kovtun:2003wp, Iqbal:2008by}. The derivation given above directly generalizes to $d$ boundary spatial dimensions, with the result \cite{Jain:2010ip, Chakrabarti:2010xy}
\be\label{eq:sq}
\sigma_Q = Z_+ \left(\frac{s}{4 \pi}\right)^{(d-2)/d} \left( \frac{sT}{\epsilon + P} \right)^2 \,.
\ee
The extra factor of the entropy for $d \neq 2$ supplies the dimensionality to the
conductivity.

With the solution (\ref{eq:aox}) at hand, we can also (re)obtain the coefficient of the delta function in (\ref{eq:finitedensityconducs}). Near the boundary
\be
a_x^{(0)}(r) \to 1 - \frac{\rho^2}{\epsilon + P} \frac{1}{r} \,.
\ee
Given that $\sigma(\omega) = G^R_{JJ}(\omega,0)/(i \omega)$, from the holographic dictionary, we obtain
\be
\sigma(\w) = \frac{\rho^2}{\epsilon + P} \left(\delta(\omega)+\frac{i}{\omega}\right) \,,
\ee
in addition to the finite part (\ref{eq:sq}).

\section{Scaling theory of \texorpdfstring{$\sigma_Q$}{$sigmaQ$}}
\label{sec:scaling}

In this section we will describe the scaling theory of $\sigma_Q$. That is, we will obtain the temperature dependence of $\sigma_Q$ in terms of various critical exponents. These will be exponents characterizing the metallic quantum critical theory to which the doped CFT flows. The analysis will be independent of holography, but will be substantiated by specific holographic examples in which $\sigma_Q$ takes the form given in equation (\ref{eq:sq}). We should emphasize that the reason it is possible to apply a scaling analysis to $\sigma_Q$ is that it is an intrinsic dissipative property of the universal low energy physics and not sensitive to irrelevant operators about that fixed point.

A general scaling framework for quantum critical metals has recently emerged from classifications
of holographic geometries. To discuss nonzero density thermodynamics and thermoelectric transport in general, three exponents are needed. The dynamical critical exponent $z$ determines the relative scaling of space and time. This is a well known
exponent \cite{subirbook} and first considered holographically in \cite{Kachru:2008yh}.
Thus
\be
[k] = - [x] = 1 \,, \qquad [\omega] = - [t] = [T] = z \,.
\ee
The hyperscaling violation exponent $\theta$ determines the anomalous scaling of the critical contribution to the
energy density and free energy (and hence also the entropy)
\be\label{eq:fscale}
[\epsilon] = [f] = z + [s] = z + d - \theta \,.
\ee
Hyperscaling violation is a well known phenomenon in statistical mechanics. Hyperscaling violation is less commonly invoked in quantum criticality, but is ubiquitous in holographic theories \cite{Charmousis:2010zz, Gouteraux:2011ce, Huijse:2011ef, Hartnoll:2011fn} and can also arises naturally in systems with Fermi surfaces \cite{Huijse:2011ef}. Finally, there can also be an anomalous scaling exponent $\Phi$ for the critical contribution to the charge density, beyond that
implied by hyperscaling violation
\be\label{eq:rhoscale}
[\rho] = d - \theta + \Phi \,.
\ee
A nonzero $\Phi$ is common in holographic models \cite{Gouteraux:2012yr,Gath:2012pg, Gouteraux:2013oca, Gouteraux:2014hca, Karch:2014mba}. It may also be necessary in order to understand some scaling properties of strange metals \cite{Hartnoll:2015sea}.
From the above formulae it follows by dimensional analysis \cite{Gouteraux:2013oca, Gouteraux:2014hca, Karch:2014mba, Hartnoll:2015sea} that the critical, universal contribution to the electrical conductivity scales as
\be\label{eq:sQscaling}
\sigma_Q \sim T^{(d-2-\theta+2 \Phi)/z} \,.
\ee

We can check whether the general expectation (\ref{eq:sQscaling}) reproduces the explicit result obtained in the previous section for a class of holographic models. In the holographic systems the total charge is not tuned to some critical value. Indeed extremal black holes typically seem to describe (`deconfined') quantum critical phases rather than quantum critical points \cite{Hartnoll:2011fn}. In particular, changing the chemical potential does not drive the system away from criticality. This suggests that the chemical potential is a marginal or irrelevant coupling in the low energy theory \cite{Hartnoll:2012rj}. The scaling of the chemical potential follows from $[f] = [\rho] + [\mu]$ and from (\ref{eq:fscale}) and (\ref{eq:rhoscale}) above. Let us consider the marginal case first. For $\mu$ to be dimensionless in the low energy scaling theory one must have
\be
\Phi = z \,. 
\ee
Thus in this case we expect
\be\label{eq:marginal}
\left. \sigma_Q \right|_\text{$\mu$ marginal} \sim T^{2 + (d-2-\theta)/z} \,.
\ee
It was noted in \cite{Hartnoll:2012rj} that indeed the ubiquitous $\sigma_Q \sim T^2$ scaling observed in holographic models with $d=2$ and $\theta = 0$ required the charge density operator to be marginal. Let us now see if this expectation (\ref{eq:marginal}) is realized more generally.

A well studied class of bulk theories are Einstein-Maxwell-dilaton theories with certain exponential potentials. Low temperature solutions to these theories exhibit hyperscaling violation, with $s \sim T^{(d-\theta)/z}$ \cite{Gubser:2009qt, Charmousis:2010zz, Iizuka:2011hg, Huijse:2011ef}, as per equation (\ref{eq:fscale}). We can use this temperature scaling directly in the holographic formula (\ref{eq:sq}) for $\sigma_Q$, together with the fact that $\epsilon + P$ will be dominated by a temperature-independent high energy contribution (this is the statement that at low temperatures $\epsilon + P \approx \mu \rho$, which is a constant). We then need to know how $Z_+$ scales with temperature. It has been found in \cite{Charmousis:2010zz, Iizuka:2011hg, Gouteraux:2011ce, Huijse:2011ef}, by solving the Einstein-Maxwell-dilaton equations of motion, that $Z_+ \sim T^{2 [(d-1)\theta/d  - d]/z}$. Therefore (\ref{eq:sq}) becomes
\be\label{eq:holo}
\left. \sigma_Q \right|_\text{holographic} \sim Z_+ s^{(d-2)/d} (s T)^2 \sim T^{2 + (d-2-\theta)/z}  \,,
\ee
in agreement with (\ref{eq:marginal}). This same scaling is also found in the frequency dependence of the low frequency, $T = 0$ optical conductivity of these theories \cite{Charmousis:2010zz}. In the appendix we discuss this match in a little more detail and also discuss the case of irrelevant charge density.

\section{Discussion}

In this paper we have discussed the physics of $\sigma_Q$, a charge transport coefficient that plays a central role in doped CFTs. Several, although not all, of the results we have discussed above have appeared previously scattered around the literature. We have placed these results within the context of current interest in identifying universal aspects of strongly interacting transport. From this perspective, the most important fact about $\sigma_Q$ is that it is the conductivity of a certain `incoherent' current that decouples from the conserved total momentum and is hence completely intrinsic to the low energy physics. With this in mind, $\sigma_Q$ could reasonably be called the `universal' or `incoherent' or `diffusive' conductivity. Perhaps the most accurate name, if a little clumsy, would be the `non-advective' conductivity. This last option captures the essential fact that it quantifies conduction of charge that is independent of the bulk fluid flow.

It is important to differentiate $\sigma_Q$ from a different interesting quantity that appears naturally in holographic formulae for the d.c.~conductivity, {\it once translation invariance is broken}. This latter quantity, which we denote as $\overline\sigma$, is the d.c.~electrical conductivity measured with the boundary condition of vanishing thermal current \cite{Donos:2014cya, Donos:2014yya}, and can reasonably be called a `pair-production' term (with the understanding that there are no quasiparticles to pair produce). The quantity $\overline \sigma$ can be shown, in certain circumstances, to be a lower bound on the d.c.~electrical conductivity of the system with respect to different ways of breaking translation invariance \cite{Lucas:2015lna, Grozdanov:2015qia}. Therefore $\overline \sigma$ can also be thought of as a `minimum conductivity'.

The distinction between $\sigma_Q$ and $\overline\sigma$ is that the former is the electrical conductivity that is independent of the bulk fluid flow in the translationally invariant limit, while the latter is the electrical conductivity that is independent of heat flow in the d.c.~limit. These two different conductivities typically have a different temperature scaling. When translation invariance is weakly broken and momentum relaxes over a long timescale $\tau$, the physics of the clean system will apply at timescales $t\ll\tau$. In particular, over these timescales the basis of currents $\{J^\text{inc},P = J^Q + \mu J \}$ diagonalizes the conductivity matrix \cite{Davison:2015bea}. However, at the longest timescales $t\gg\tau$ that control the d.c.~conductivities, there is a reorganisation of transport in a large class of holographic theories \cite{Donos:2015gia, Lucas:2015lna, Banks:2015wha}: $J^Q$ takes the role of $P$ \cite{Blake:2015epa, Blake:2015hxa} and the conductivity matrix is now diagonal in the basis $\{J^\text{inc},J^Q\}$. Note, however, that both $\sigma_Q$ and $\overline\sigma$ can be defined in the clean theory. With strong breaking of translation invariance, momentum is generically no longer a privileged operator and should not be expected to play a significant role in transport.

This work has been in the framework of a CFT deformed by a charge density. However, in the limit in which they are large, non-advective conductivities can be defined and discussed in complete generality -- without Lorentz invariance -- using the memory matrix formalism \cite{Lucas:2015pxa}. In general there are three such conductivities, denoted $\sigma_Q$, $\alpha_Q$ and $\bar{\kappa}_Q$ in \cite{Lucas:2015pxa}, due to the existence of two independent incoherent currents. In addition to the incoherent electrical current $J^\text{inc}$ of (\ref{eq:generalincoherentcurrent}), there is an analogous incoherent heat current, given by replacing $J$ with $J^Q$ in (\ref{eq:generalincoherentcurrent}). In a CFT deformed by a charge density, these currents are equivalent (because momentum is equal to the energy current) and thus there is only one non-advective conductivity $\sigma_Q$ in that case.

Finally, because $\sigma_Q$ is an intrinsic, incoherent conductivity associated to a diffusive process in a metal, it may be a natural quantity to attempt to bound in the spirit of \cite{Hartnoll:2014lpa}. Such a bound may be relatively simple to explore, because $\sigma_Q$ is defined in translation-invariant (but finite density) systems and is physically similar to the shear viscosity \cite{Kovtun:2004de}.

\section*{Acknowledgements}

It is a pleasure to acknowledge Jan Zaanen and Nabil Iqbal for stimulating discussions. We also thank the KITP, KITPC and the GGI for hospitality, and the INFN for partial support, during the completion of this work. The work of R.D. is supported by a VIDI grant from NWO, the Netherlands Organisation for Scientific Research. The work of B.G. is supported by the Marie Curie International Outgoing Fellowship nr 624054 within the 7th European Community Framework Programme FP7/2007-2013. The research of S.A.H. is partially supported by an Early Career Award from the US Department of Energy.

\appendix

\section{Scaling analysis of Einstein-Maxwell-dilaton solutions}

The match between the scaling expectation (\ref{eq:marginal}) and the holographic answer (\ref{eq:holo}) indicates that the charge density operator has become marginal in the IR fixed point of the Einstein-Maxwell-dilaton spacetimes. It is instructive to see directly how this works out and, in doing so, also discuss the case of irrelevant charge density. The IR fixed point is described by an extremal (i.e. $T = 0$) near horizon geometry of the form\footnote{The reader should beware that here we are taking $r \to 0$ in the IR -- consistent with the coordinates used in the main text -- whereas many other discussions use an inverted coordinate in which $r \to \infty$ towards the IR. For simplicity, we restrict ourselves to the parameter space $\theta<d$, $z>1$ where the location of the IR in the metric (\ref{eq:ds2}) is unambiguous.}
\be\label{eq:ds2}
ds^2 = \frac{1}{r^{2 \theta/d}} \left(r^{2z} dt^2 + \frac{dr^2}{r^2} + r^2 dx^2_d \right) \,.
\ee
This metric (\ref{eq:ds2}) geometrizes the critical exponents $z$ and $\theta$.
Furthermore, in these solutions the Maxwell field takes the general form
\be\label{eq:zeta}
A_t = r^{z-\zeta} \,.
\ee
This is the definition of the exponent $\zeta$ appearing in \cite{Gouteraux:2012yr, Gouteraux:2013oca, Gouteraux:2014hca, Donos:2014oha,Kiritsis:2015oxa,Kiritsis:2015yna}, which we now wish to relate to the anomalous dimension $\Phi$ of the charge density operator.

We now show that, depending on a choice of quantization, there are two possibilities for the behavior of the Maxwell field as a function of the anomalous dimension of the charge density operator:
\be\label{eq:Att}
(\text{I}): \; A_t = \frac{1}{r^{d_\text{eff.} +2 (\Phi - z)} } \qquad \text{or} \qquad (\text{II}):\; A_t = r^{d_\text{eff.}  + 2 (\Phi - z)}  \,.
\ee
Here the effective number of spacetime dimensions $d_\text{eff.} = d + z - \theta$. In particular, the dimensions of operators $\ocal$ and their corresponding sources $g$ obey $[g] + [\ocal] = d_\text{eff.}$. We are presently considering the case $\ocal = \rho$ and $g = \mu$. The electrostatic potential $A_t$ will have two independent modes, one given in (\ref{eq:Att}) and the other being $r^0$ (by gauge invariance, this is always a solution). In holography, one of these modes must correspond to the expectation value $\langle \ocal \rangle$ and the other to the source $g$ \cite{Hartnoll:2009sz}. In particular, the difference in the two exponents must be $\pm (2 [\ocal] - d_\text{eff.})$. The ambiguity in the sign depends upon which mode we consider to be the source and which to be the expectation value. The expressions (\ref{eq:Att}) are then obtained by recalling that $[\ocal] = [\rho] = d_\text{eff.} + \Phi - z$.

For marginality, $\Phi = z$. The solutions in the literature  \cite{Charmousis:2010zz, Iizuka:2011hg, Gouteraux:2011ce, Huijse:2011ef} have $\zeta=\theta-d$ in the Maxwell potential (\ref{eq:zeta}). Equating (\ref{eq:zeta}) and (\ref{eq:Att}) with these values we can conclude that the solutions can indeed be interpreted as being marginal if we use the quantization leading to case (II) in (\ref{eq:Att}).
This corresponds to the natural quantization of the Maxwell field (assuming $d_\text{eff.} > 0$) in this setting, in which the larger mode towards the UV boundary of the IR geometry, $r \to \infty$, is treated as the source while the subleading $r^0$ mode is the response. Note that this is different from the more familiar quantization (without an anomalous dimension for the charge density) in which the constant $r^0$ mode is the source.

In the case of an irrelevant charge density, one expects Lorentz invariance to be restored and hence $z=1$ at the low energy fixed point. The exponent $\Phi$ is then not fixed but constrained to satisfy $1 < \Phi$. Such solutions have also been found in Einstein-Maxwell-dilaton theory, in a different regime of parameter space to those above \cite{Gouteraux:2014hca, Donos:2014oha}. For these solutions, if we match the behavior of $A_t$ to (\ref{eq:Att}) to extract $\Phi$, we find
\be\label{eq:PhitoZeta}
(\text{I}): \; 2\Phi=\zeta+\theta-d \qquad \text{or} \qquad (\text{II}):\; 2\Phi=2-d+\theta-\zeta  \,.
\ee
The scaling result (\ref{eq:sQscaling}) then yields the predicted temperature scalings 
\be
\label{eq:scalingzeta}
(\text{I}): \; \sigma_Q \sim T^{\zeta-2}\qquad \text{or} \qquad (\text{II}):\; \sigma_Q\sim T^{-\zeta}  \,.
\ee
On the other hand, from the holographic formula (\ref{eq:sq}) we get
\be\label{eq:irr}
\sigma_Q\sim T^{2(d-\theta)+\zeta}\,,\qquad \epsilon+P\sim T^0\,,\qquad Z_+\sim T^{\zeta-d+\theta-\frac2d\theta} \,,
 \ee 
 where we have assumed that $\epsilon+P$ goes to a constant at $T=0$ and the scaling of $Z_+$ is fixed by the solution. We see that the actual temperature dependence (\ref{eq:irr}) of $\sigma_Q$ from the holographic solution does not match the scaling result (\ref{eq:scalingzeta}) for either choice of quantization. We suspect that this is because the irrelevance of the operator introduces extra dimensionful scales into the IR solutions. We can note, however, that the scaling for the choice of quantization (II) in (\ref{eq:scalingzeta}) matches the $T=0$, low frequency scalings derived in \cite{Charmousis:2010zz,Kiritsis:2015yna}. On the other hand, the scaling for the choice of quantization (I) matches the temperature scaling of the DC conductivity derived in \cite{Gouteraux:2014hca} for $\overline \sigma$ in the presence of momentum relaxation.

\end{document}